\documentclass[preprint,12pt]{revtex4}
\usepackage{dcolumn}
\usepackage{amsmath,amsfonts,amssymb} 
\usepackage{graphicx}
\usepackage[usenames]{color}

\def\a{\alpha} 
\def\b{\beta}
\def\g{\gamma}
\def\G{\Gamma}
\def\d{\delta}

\def\l{\lambda}

\definecolor{myblue}{rgb}{0,0.1,0.6}
\definecolor{myorange}{cmyk}{0,0.6,0.8,0} 
\newcommand{\red}[1]{\textcolor{red}{#1}}

\begin{document}

\title{Modeling Truncated Hemoglobin vibrational dynamics}

\author{Luca Marsella\footnote{E-mail: marsella@sissa.it}}
\affiliation{International School for Advanced Studies (S.I.S.S.A.), 
via Beirut 2-4 34014 Trieste, Italy}

\date{\today}

\begin{abstract}
We present a  study on the near equilibrium dynamics of two small
proteins in the family of truncated hemoglobins, 
developed under the framework of a Gaussian network approach.
Effective beta carbon atoms are taken into account besides $C^{\a}$s
for all residues but glycines in the coarse-graining procedure, 
without leading to an increase in the degrees of freedom ($\b$Gaussian Model).
Normalized covariance matrix and deformation along slowest modes with
collective character are analyzed, pointing out anti-correlations 
between functionally relevant sites for the proteins under study.
In particular we underline the functional motions of an extended 
tunnel-cavity system running inside the protein matrix,
which provide a pathway for small ligands 
binding with the iron in the heme group. 
We give a rough estimate of the order of magnitude 
of the relaxation times of the slowest two
overdamped modes and compare results with previous 
studies on globins.\\

\end{abstract}

\maketitle

\section{Introduction}

Several studies in the past decades have shown the validity of the
normal modes approach to extract useful information on the
large-scale functional movements of proteins near 
their native state conformation~\cite{noguti, go, brooks, levitt-sander,levitt-stern}.

Molecular dynamics simulations of biomolecules performed 
using detailed all atoms potentials yield lots of
information regarding large amplitude, concerted displacements 
of atoms~\cite{hayward1}. However these can be simply obtained within the
harmonic approximation and from the analysis of the hessian matrix:
in fact only low-frequency modes
provide the major part of the norm for those global
motions, whereas the fastest modes account for only spatially localized
fluctuations~\cite{hayward2, amadei}.

It has become customary to project the dynamical trajectories
of the atoms in the molecules
onto normal mode axes~\cite{horiuchi}; thus one is brought to interpret the functional, 
large amplitude motions of biological relevance for
proteins as superpositions of independent harmonic modes of oscillations of a
network of atoms.

A pioneering work developed by Tirion~\cite{tirion} paved the
way for extremely simplified Normal Mode Analysis (NMA):
detailed harmonic potentials are replaced by a single-parameter, 
spring-like potential between atoms found to be in contact in the native configuration.

Despite the extreme simplicity of this approach, 
the good agreement obtained with atomic mean square displacements 
of molecular dynamics simulations~\cite{tirion} opened the
possibility for further studies, within the same
approach~\red{\cite{haliloglu, bahar1, bahar2, doruker, atilgan, jernigan, tama1, tama2}}: 
a good level of consistency with more accurate analyses is achieved
even treating proteins under coarse-grained 
schemes, \red{as recently shown by Bahar and 
co-workers~\cite{haliloglu, bahar1, bahar2, doruker, atilgan, jernigan},
who developed simple yet useful models to explore the collective
motions of proteins, profitably adopted 
by other groups~\cite{mich, neri}.
}

In the present study two structures recently
solved~\cite{pesce} are addressed, 
which belong to the family of truncated 
hemoglobins (trHbs), small heme proteins
widely distributed in bacteria, protozoa and plants,
forming a distinct group within the hemoglobin
super-family~\cite{couture, yeh, bolognesi}.

Though having a simpler structure than the traditional globin fold,
they still preserve the respiratory function, providing transport and
storage of oxygen molecules.
Furthermore they have been proposed to be involved also in other biological
functions, such as protection against reactive nitrogen species, 
photosynthesis or to act as terminal oxidases~\cite{potts, thorst, couture94, couture}.

The low complexity of trHbs structure, compared to normal globin folds, 
might help the comprehension of the mechanisms used by these shorter molecules 
to bind small ligands to the heme iron atom (e.g.: $O_2$, their main target,
and $CO$, to which heme has a high affinity).

In particular, the presence of an apolar cavity system extending
throughout the protein matrix of truncated hemoglobin from
\emph{Mycobacterium Tuberculosis} and homologous structures 
has been recently noticed~\cite{milani1, bolognesi}: 
this tunnel connects the heme distal pocket to the protein surface,
and may thus allow an efficient diffusion path for oxygen and other small
molecules to the iron atom (fig.\ref{fig:trHb}).

The role of protein cavities has been deeply investigated
in myoglobin~(see \cite{brunori, schotte, bolognesi} and references therein), 
both theoretically using computer simulations and experimentally
suggesting pathways for ligands migration switched by a small number
of substates, which can be allosterically converted to the stable
conformations~\cite{teeter}.

These issues are investigated here from a novel point of view, through a simple 
coarse-grained scheme in the spirit of the Gaussian chain models, 
with a twofold goal: understanding the mechanical processes 
involved in the functional movements of these key proteins and
taking advantage of this new Gaussian framework, 
computationally fast and conceptually simple.

\section{Structural characterization}

\begin{figure}[htbp]
\begin{center}
\includegraphics[width=\textwidth]{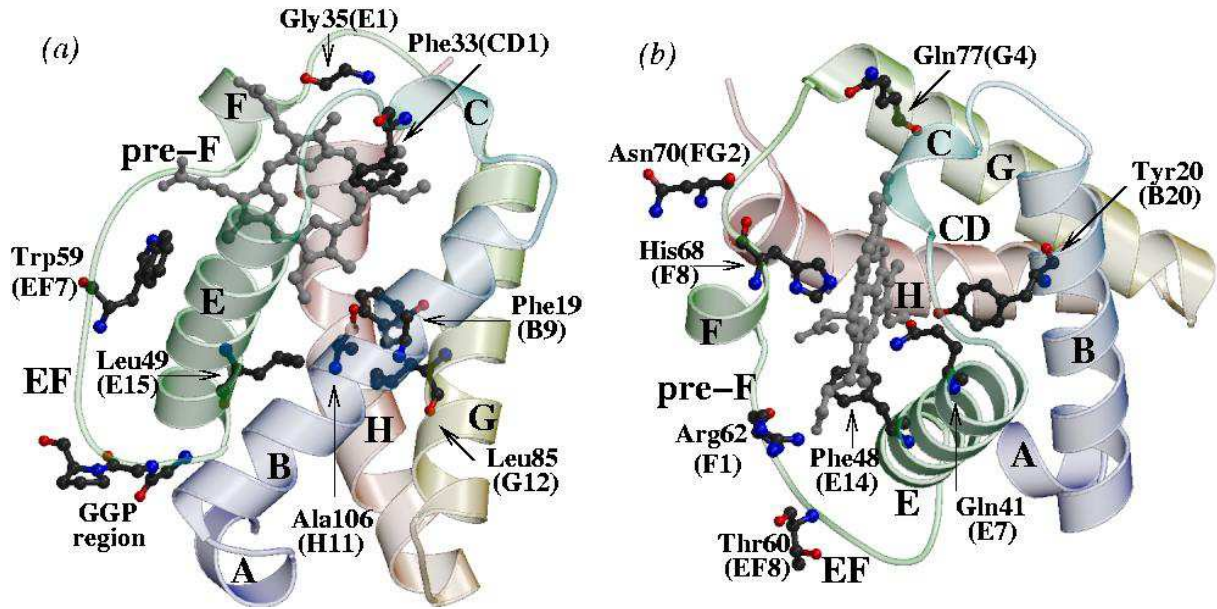}
\end{center}
\caption{Truncated hemoglobin fold from \emph{Paramecium Caudatum}:
helices, coils and the main residues described in the text 
are labeled according to the standard nomenclature for globins.
The two-over-two helical structure enclosing the
heme group is clearly visible: (a) side view, (b) top view.
Figure drawn using Molscript~\cite{mol} and Raster3d~\cite{render}.}
\label{fig:trHb}
\end{figure}

The structures addressed in the present study are two truncated
hemoglobins from the ciliated protozoan
\emph{Paramecium Caudatum} (PtrHb, pdb id: 1dlw) 
and the green unicellular alga \emph{Chlamydomonas Eugametos} (CtrHb, pdb id:
1dly), solved at 1.54 and 1.80~\AA \, resolution respectively, by
Pesce et al.~\cite{pesce}.

Similarly to the other proteins belonging to the trHb family, 
they display low sequence identity with hemoglobins
from vertebrate and non vertebrate.
This is smaller than 15\% for PtrHb and CtrHb, due to substantial
residue deletions at either N- or C- termini and in the C and D
helical region of the globin fold~\cite{pesce}.

More than 70\% of the residues in the two structures belongs to
helices, mainly of type $\a$ 
(above 67\% in both proteins: only the short helix C is of type
$3_{10}$): this is a typical feature of the globin fold, which leads to
guess a primary role of helices in the functional motions of these
proteins, as well as in myoglobin and hemoglobin.
Nonetheless several structural differences make truncated
hemoglobins fall in a distinct group in the 
hemoglobin superfamily~\cite{bolognesi, pesce}.

Helices in the globin fold are traditionally indexed through capital
letters A, B, C, D, E, F, G and H, while loops between them are named
according to the nearby helices, and residues are numbered
sequentially with each unit~\cite{perutz}.

The structures taken in consideration here 
reveal the so called ``two over two $\a$ helical sandwich''(fig.~\ref{fig:trHb}), 
in place of the classical ``three over three'' observed in the globin fold~\cite{holm}:
in fact helix D is absent, while N-terminal A helix and proximal F
helix are drastically reduced to only one turn.

A structure-based sequence alignment of PtrHb, CtrHb and other
truncated hemoglobins with sperm whale Myoglobin, reported in
\cite{pesce}, shows the strongly conserved residues among proteins in
the trHbs family, mainly of three types:
\begin{description}
  \item[glycine rich motifs]
    especially at helices termini, which enhance
    structural flexibility (Gly-Gly motifs at the beginning of the AB
    and EF regions, and Gly-Arg/Lys in the pre-F region~\cite{pesce});
    
  \item[hydrophobic residues]
    on heme distal and proximal sides, which
    play the main role of shielding heme from solvent molecules, 
    in order to prevent iron oxidation~;

  \item[heme binding residues]
    stabilizing the porphyrin ring in the heme
    pocket; one particularly relevant is the proximal histidine,
    His 68, localized on helix F.

\end{description}

Strongly conserved residues on the distal side responsible for
the shielding of the heme pocket from the solvent are mainly localized
on helices B and E, as well as in the CD and EF loops: hydrophobic
residues Phe A12, B9, CD1, E14 and Trp EF7, with their side chains
pointing to the inner part of the molecule; Tyr B10, Gln E7, with
side chains responsible for the stabilization of the ligand 
bound to heme~\cite{pesce, bolognesi}. 

The hydrophobic residues identified in~\cite{pesce, milani1}
as the ones defining a cavity inside the molecule, linking 
the solvent exposed surface of the proteins to the heme group 
are positioned on the distal side.
They are mainly localized on helices A (at the opening of the tunnel on the
surface), B, E (limiting the distal side) and G.

On the proximal side of the heme pocket one finds the proximal
histidine, in a strongly conserved position within hemoglobin (Hb) 
and trHb families: the imidazole ring of histidine allows it to act as either a proton
donor or acceptor at physiological pH. 
In hemoglobins is essential its ability to buffer the $H^+$ ions 
from carbonic acid ionization in red blood cells, 
allowing the molecule to exchange $O_2$ and
$CO_2$ respectively at the tissues and at the lungs~\cite{biochem}.

It will be shown how the small $\a$ helix F, which contains the proximal
histidine F8, can play a leading role as a reference position 
for elucidating the functional motions of the protein regions 
around the heme pocket.

\section{Theory}
\color{red}
The model adopted in this study is the Beta Gaussian Model 
($\b$GM) presented in \cite{mich}, a single parameter model 
apt to describe small amplitude fluctuations of residues 
around their native-state equilibrium:
the model is based on the Gaussian Network Model (GNM) and on the 
Anisotropic Gaussian Model (ANM), 
which have been successfully exploited in previous
studies on the functional motions of
proteins~\cite{haliloglu, bahar1, bahar2, doruker, atilgan, jernigan},
under the framework of the single parameter model introduced by Tirion~\cite{tirion}. 

Only alpha and beta carbon atoms ($C^{\a}$, $C^{\b}$)
are treated: rather than the actual $C^{\b}$, the latter is 
an effective centroid accounting for the directionality of the side
chain, built for all residues but glycines and
terminus ones; its position is determined by 
the coordinates of neighbouring $\a$ carbons~\cite{park,mich},
according to the following relation:

\begin{equation} \label{eqn:cb}
  {\bf r}^{\b}_i = {\bf r}^{\a}_i + 
  3 \: \frac{ 2 {\bf r}^{\a}_i - {\bf r}^{\a}_{i+1} - {\bf r}^{\a}_{i-1}}
  { |2 {\bf r}^{\a}_i - {\bf r}^{\a}_{i+1} - {\bf r}^{\a}_{i-1}|}  
\end{equation}

where the vectors ${\bf r}^{\a}_i$ and ${\bf r}^{\b}_i$ hold the native
coordinates (in~Angstroms) of the $\a$ carbon atom and of the effective $\b$ centroid
which belong to the $i$th residue.
Expanding the displacement of the $C^{\b}$ from the equilibrium 
to leading order in the displacements of the $C^{\a}$s one gets:
\begin{equation} \label{eqn:deltacb}
  \d{\bf r}^{\b}_i \sim
  3 \: \frac{ 2 \d{\bf r}^{\a}_i - \d{\bf r}^{\a}_{i+1} - \d{\bf r}^{\a}_{i-1}}
  { |2 {\bf r}^{\a}_i - {\bf r}^{\a}_{i+1} - {\bf r}^{\a}_{i-1}|}  
\end{equation}

The hamiltonian of the system depends quadratically on the deviations
of the $C^{\a}$ and $C^{\b}$ from their native positions, 
assumed to be the energy minimum in the configuration space 
(thus neglecting crystal effects on X-ray structures): 
the displacements of protein's atoms from the equilibrium position 
are supposed to be small enough to justify 
this approximation~\cite{noguti, doruker, horiuchi}.

The hamiltonian includes interactions between $\a$ and $\b$
carbons lying within a cut-off distance $r_{c}$, above which no pairwise
interaction is allowed, as well as an effective interaction
accounting for the strength of the peptide bond for nearest-neighbouring $C^{\a}$s:

\begin{equation} \label{eqn:hamiltonian}
  \mathcal{H} = \mathcal{H}^{peptide}+\mathcal{H}^{\a\a}+\mathcal{H}^{\a\b}+\mathcal{H}^{\b\b}
\end{equation}

where
\begin{eqnarray} \label{eqn:betagm}
 \mathcal{H}^{peptide} & = & \frac{\g_p}{2} \sum_{i} \sum_{\mu, \nu}
 \mathcal{M}_{i \: i+1}^{\mu \nu}(\a,\a)\:\d r^{\a}_{i,\mu} \d
 r^{\a}_{i+1,\nu} \nonumber\\
 \mathcal{H}^{xy} & = & \frac{\g_{xy}}{2} 
 \left(1 - \frac{\d_{xy}}{2} \right) \sum_{i,j} \sum_{\mu,\nu}
 \mathcal{M}_{ij}^{\mu \nu}(x,y)\:\d r^x_{i,\mu}\d r^y_{j,\nu}
\end{eqnarray}

\begin{description}
\item
$\g_p$ is the elastic constant accounting for
the relative strength of the effective peptidic interaction 
between nearest-neighbouring $\a$ carbons;

\item
$\g_{xy}$ is the elastic constant for the 
contact interaction between carbon atoms of type $x$ and $y$ 
($x,y \in \{\a,\b\}$);

\item
$\d_{xy}$ is Kronecker delta to avoid double counting of the
interactions between atoms of the same type;

\item
$\d {\bf r}^x_i$ is the displacement from the native position 
of the carbon atom of type $x$ that belongs to the $i$th residue
($\mu$ and $\nu$ are the indexes of the Cartesian components);

\item
$\mathcal{M}_{ij}(x,y) \: (i \neq j)$
is a $(3 \times 3)$ matrix, the off-diagonal 
super-element of the hessian matrix for the 
interaction between atoms of type $x$ and $y$ which belong to residues
$i$ and $j$:
\begin{equation} \label{eqn:off-diagonal}
  \mathcal{M}_{ij}^{\mu \nu}(x,y) = 
  \G_{ij}^{xy} \:
  \frac{r^{xy}_{ij,\mu} \: r^{xy}_{ij,\nu}}
       {{\bf r}^{xy}_{ij} \cdot {\bf r}^{xy}_{ij}}
\end{equation}
where $\G^{xy}_{ij} \: (i \neq j)$ is equal to 1 if the native separation 
of the corresponding atoms lies below the cut-off radius $r_c$, 0 
otherwise; ${\bf r}^{xy}_{ij} = {\bf r}^{x}_{i} - {\bf r}^{y}_{j}$ 
is the vector of native separation 
of atoms of type $x$ and $y$ that belong to residues $i$ and $j$
respectively. Entries of diagonal super-elements 
are built according to the relation:
\begin{equation} \label{eqn:diagonal}
  \mathcal{M}_{ii}^{\mu \nu}(x,y) = - \sum_{j \neq i} {M}_{ij}^{\mu \nu}(x,y)
\end{equation}
\end{description}

Since the position of the effective $C^{\b}$ and its displacement
from equilibrium are fully determined by $\a$ carbons coordinates
( equations (\ref{eqn:cb}) and (\ref{eqn:deltacb}) ),
by substitution of (\ref{eqn:cb}) and (\ref{eqn:deltacb}) in (\ref{eqn:betagm})  
one is left with an effective hamiltonian $\tilde{\mathcal{H}}$
which depends quadratically on $C^{\a}$ 
displacements from native state~\cite{mich}
(the index of atom type will be therefore 
dropped in the following equations for simplicity):
\begin{equation} \label{eqn:effective}
\tilde{\mathcal{H}} = 
\frac{\g}{2} \sum_{ij} \sum_{\mu \nu} 
\tilde{\mathcal{M}}_{ij}^{\mu\nu} \d r_{i,\mu} \: \d r_{j,\nu}
\end{equation}
where $\g_p$ and $\g_{xy} \: (x,y \in \{\a,\b\})$ 
have been incorporated in $\tilde{\mathcal{M}_{ij}}$, 
expressed in units of the reference elastic constant $\g$.

\color{black}
Time dependent two-point correlation functions can be 
calculated within a Langevin dynamics leading 
to equilibrium with the Boltzmann factor 
$e^{-\b\:\tilde{\mathcal{H}}}$~\cite{bahar1}.
In the overdamped regime with the viscous damping factor $f$, the same
for all residues~\cite{haliloglu}, and white noise $\eta_i(t)$ , the
Langevin equation for our system is~\cite{doi}:
\begin{equation} \label{eqn:langevin}
f \frac{d}{dt} {\d r_{i,\mu}}(t) + 
\g \sum_{j,\nu} \tilde{\mathcal{M}}_{ij}^{\mu\nu} \: \d r_{j,\nu}(t) = \eta_i(t)
\end{equation}
 
One can easily get from equation (\ref{eqn:langevin})
the time dependence of cross correlations between
couples of $C^\a$s (the so-called ``reduced'' cross-correlations):
\begin{equation} \label{eqn:time-correlation}
\left<\d {\bf r}_{i} (t)\cdot \d {\bf r}_{j} (0)\right> =
\frac{k_{B}T}{\g} \sum_{k}\frac{1}{\l_k}
\left({\bf a}_{ik} \cdot {\bf a}_{jk}\right)\;
e^{-\l_k \: \frac{t}{\tau}}
\end{equation}

$\tau = \frac{f}{\g}$ is the reference relaxation time, corresponding to an
overdamped spring of elastic constant $\g$ in a dissipative medium of
friction $f$; $\l_k$ are non zero eigenvalues of
$\tilde{\mathcal{M}}$ and ${\bf a}_{k}$ the corresponding
eigenvectors.

Theoretical B-factors (measured in \AA$^{2}$) 
are obtained from the diagonal elements of the
reduced covariance matrix 
(i.e. from the mean square fluctuations of
$C^\a$s around native-state equilibrium, 
after thermal equilibrium has been reached), 
through the relation:
\begin{equation} \label{eqn:B-factors}
B_i = \frac{8\pi^2}{3} \frac{kT}{\g}
\left< \d{\bf r}_i \cdot \d{\bf r}_i \right>
\end{equation}

Equation (\ref{eqn:B-factors}) will be used to fit 
the experimental B-factors and get
an estimate of the elastic constant $\g$.

\subsection{Tuning model parameters}

In order to obtain reliable data for the structures under study, we
compare theoretical and experimental results using the ranking
correlation between the two data sets as a guideline to tune model
parameters to their optimal values.

\color{red}
ANM was applied on the structures as well as $\b$GM:
in the case of the trHb, the theoretical temperature factors 
obtained with the $\b$GM showed a higher value of 
Kendall's non parametric $\tau$~\cite{numrec}(see below)
against the experimental ones ($\tau = 0.45$ for ANM with $r_c = 13.0$ \AA\:, 
$\tau = 0.57$ for $\b$GM with $r_c = 7.0$ \AA\:, in the case of 1DLW).

ANM works very well for bigger complexes, while for smaller proteins
more details are required: the reason for the better agreement 
obtained by the $\b$GM is to be found in the presence of the $\b$
centroids, which considerably increases the number of 
pairwise interactions and takes into account the directionality 
of the side chains is in the contact map of $\a$ carbons.

As a consequence, the $\b$GM needs a lower and more realistic 
cut-off radius $r_c$ to reproduce experimental B-factors and molecular dynamics
data, in comparison to those used by ANM~\cite{atilgan},
as already remarked in a previous study~\cite{mich}, even with small
proteins like trHbs: hence the choice to use the $\b$GM in the present work.

Here in particular, the best agreement between theory and
experiment was found using a cut-off of 7.0~\AA.
\color{black}
This choice is imposed by the difference in compactness between helical
regions and coils, and it is critical 
in order to keep the density of effective contact interactions 
at the coarse-grained level comparable to the all atoms one.

Larger cut-offs cause contact density to be overestimated for the helical regions,
leading to smaller values of B-factors, with respect to the
experiment: the consequence is a marked difference between
flexible and solvent exposed parts of the protein, compared to the
less flexible and buried parts (fig.~\ref{fig:bfactors}).

A key point was the tuning of $\g_p$, the ratio between the effective peptide bond
and the $\a\a$ interaction: it accounts for the relative stiffness of the
covalent bonds along the backbone as opposed to the weaker contact
interactions between $C^\a$ pairs.

Summarizing the values for the parameters used in
the calculation for both structures,
$r_c = 7.0 \mbox{ \AA}$, $\g_p = 2.0$, $\g_{\a\a} = \g_{\a\b} = \g_{\b\b} =
1.0$ (the last ones are in units of $\g$).
The value of the elastic constant $\g$ will be determined later, 
fitting the results of the model with the available experimental data.

\subsection{Temperature factors and heme modeling}

Truncated hemoglobins are heme proteins, the heme group being 
the active site of the molecule: there oxygen and carbon oxide bind
to the sixth coordination position of the iron atom, which  
lies at the center of the tetrapyrrole ring and is bound to the 
imidazole ring of the proximal histidine F8 at the fifth coordination
site (His 68, eighth residue of helix F in sperm whale myoglobin and in 
vertebrate hemoglobins, where nomenclature ``F8'' comes from~\cite{stryer}).

\begin{figure}[htbp!]
\includegraphics[width=0.8\textwidth]{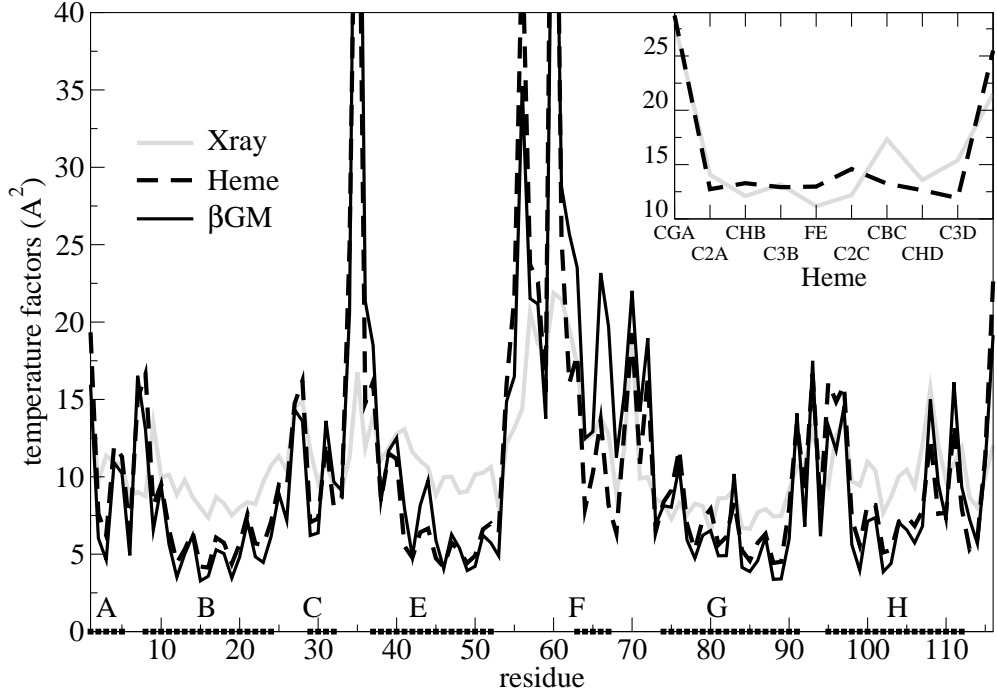}
\caption{Theoretical (black) versus experimental (gray) X-ray B-factors for $\a$
  carbons in PtrHb, related through equation~\ref{eqn:B-factors}. 
  Theoretical B-factors including coarse-grained heme group 
  are shown for comparison (dashed). Helical segments have been marked on residues axis.
  The inset shows theoretical versus experimental B-factors for
  coarse-grained heme, \red{with pdb names of iron and carbons 
    included in the coarse-graining.}}
\label{fig:bfactors}
\end{figure}

Figure \ref{fig:bfactors} shows the plot of the $\a$ carbon atoms
B-factors of the X-ray structure of truncated hemoglobin
in \emph{Paramecium Caudatum} and their corresponding mean square 
displacements derived from the $\b$GM: 
\red{most mobile regions are loops and turns between helices, 
which on the contrary display smaller fluctuations, 
in agreement with the results of an NMA study performed 
on deoxymyoglobin (Mb)~\cite{senoy1, senoy2}.}

The significance of the correlation between experimental and theoretical
values is deduced from Kendall's non parametric $\tau$~\cite{numrec}.
Since one does not know \emph{a priori} the probability distribution of
the experimental B-factors, a significance for the agreement between 
the two data sets cannot be computed from the value of the linear 
correlation coefficient.
On the other hand, the rank correlation given by $\tau$ is independent from
the distribution.
Kendall's $\tau$ for \emph{PtrHb} is 0.56 (0.52 with heme), for
\emph{CtrHb} is 0.37 (0.40 with heme), and $P_{null}(\tau) <\;10^{-9}$
in all cases
($P_{null}(\tau)$ is the probability for two random sets of data to
have a value of $\tau$ bigger than the one found between B-factors
predicted by the model and calculated from X-ray structure. 
The number of residues is 116 for PtrHb and 121 for CtrHb). 

\red{
The coarse-graining on the heme group includes the iron atom
and nine carbons of the porphyrin ring \red{(whose names are reported on
the x axis of the inset in fig.~\ref{fig:bfactors})}, 
chosen in order to keep the number of contacts in the modeled system
comparable to the number of heme native contacts with nearby residues, 
thus avoiding to have a loosely connected group as an artifact of the
coarse-graining procedure.
}

Insertion of heme brings only one relevant change to the temperature factors
plot (fig.~\ref{fig:bfactors}): helix F has displacements from
equilibrium considerably damped, as it was expected, being bound to
the iron atom. A reduction in the fluctuations is shown also 
by the loops between helices C and F, to a lesser extent than in helix F.
\red{
The protein part of the reduced covariance matrix obtained
including the coarse-grained heme was compared with the covariance matrix
computed without modeling the tetrapyrrole ring.
The two show a Kendall's parametric correlation $\tau \sim 0.81$ 
over more than thirteen thousands of points, which stands for a
remarkable agreement between them: 
the coarse-grained heme in fact anticorrelates
with the same parts of the protein as helix F, even if more weakly
(data not shown).
This is not surprising, since the iron atom and the proximal histidine
F8 are in direct contact, so the motion of the heme group 
will be strongly correlated with the one of the F helix, 
following the proximal side in its deviations from native-state
equilibrium: the inclusion of few more atoms under the coarse-grained
scheme adopted here do not seem to significantly modify the
correlations.
The mechanical response of the protein upon binding of
ligands on the iron atom is given by the properties of
the network of backbone atoms: 
thus a good agreement with known behavior of globins
may be achieved using gaussian models even without considering
heme groups in the coarse-graining procedure~\cite{xu}.
}

The $\b$GM heme B-factors plot is in substantial
agreement with the experimental B-factors for heme 
(fig.~\ref{fig:bfactors}, smaller plot).
In fact the heme pocket is entirely surrounded by non polar
residues: one of the main purposes of the distal region is 
to screen the heme group from solvent interaction, 
in order to avoid iron oxidation~\cite{stryer}.

The results for Kendall's ranking lie in the typical range 
of gaussian models~\cite{mich}, even with terminal residues 
included, and the confidence of the correlation is no doubt 
statistically significant: 
still there are some regions of the protein whose
fluctuations are not well reproduced by the model,
as shown in the plot of temperature factors 
(fig.~\ref{fig:bfactors}).

The model overestimates interactions between $\a$ and $\b$ carbons belonging 
to secondary structures, resulting in local deviations from the
density of the all atom picture. 
Hence displacements of residues belonging to helical regions are
underestimated, since these are the most compact parts of the protein
and it produces deviations in the profile of B-factors, 
whose values depend both on the assembly of secondary
motifs~\cite{lattanzi} and on the local packing density~\cite{halle}.

Furthermore, electrostatics and solvent exposure
for different residues are not taken into account by the simple
approach of the model:
electrostatic interactions localized on helices
may modify the magnitude of the driving forces producing 
larger displacements from native state than expected.

POPS program (Parameter OPtimized Surface~\cite{pops}) 
has been used to calculate the solvent accessible surface area per
residue for PtrHb: the most exposed residues are the ones displaying 
the greatest average displacements from the native structure, 
as it was expected~(figg.~\ref{fig:bfactors},~\ref{fig:12}).
These small residues (Gly 35, the GGP region - Gly54, Gly55, Pro56 - 
Thr60, Gly61), located in loops CD, EF and to the pre-F region,
allow larger flexibility to the polypeptide chain (glycines especially)
and the bigger fluctuations predicted by the model are due to their 
diminished connectivity as well, being the most exposed to the solvent.
This was expected, since the model totally neglects solvent exposure.

The simplified approach used here 
shows a remarkably better agreement with experiment, 
for buried regions, where the connectivity of atoms 
is greater and the solvent plays a minor role.

\section{Results and Discussion} 

In order to identify the relevant motions of the protein
the reduced covariance matrix plot (figure~\ref{fig:covmat}) 
of PtrHb is inspected (PtrHb will be the main target of the following discussion, 
the same considerations holding for CtrHb as well), 
normalized as follows: 

\begin{figure}[htbp]
\begin{center}
\includegraphics[width=0.5\textwidth]{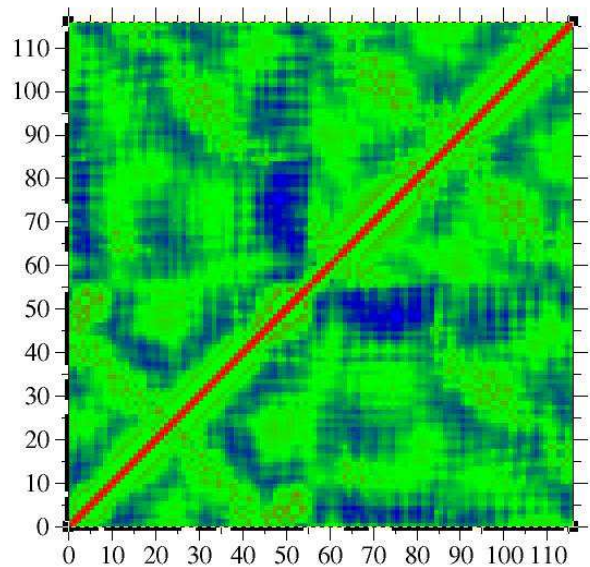}
\end{center}
\caption{
  Normalized covariance matrix: trivial correlations due to
  contacts have been put to 0 (green); diagonal elements are equal to
  1 (red); anti correlation range from 0 to the minimum value found,
  for Gln41 (E7) and His68 (F8), lower then -0.35 (blue). 
  Helical regions have been highlighted.}
\label{fig:covmat}
\end{figure}

\begin{equation} \label{eqn:ncovmat}
c_{ij} =
\frac{\left<\d {\bf r}_{i}\cdot \d {\bf r}_{j}\right>}
{\sqrt{\left< \d{\bf r}_i \cdot \d{\bf r}_i \right> 
\left< \d{\bf r}_j \cdot \d{\bf r}_j \right>}}
\end{equation}

Normalization is generally performed in order to allow a direct comparison between
the cross-correlations predicted by the model and the ones
obtained in computer simulations, e.g. from molecular dynamics,
provided equilibration has been reached~\cite{hess}.

From the reduced normalized covariance matrix one is able to 
extract non trivial informations on the collective motions of the
protein under study: these generally involve the regions of
the molecule that show negative correlations. 

Indeed it turns out that spatially closed parts of the molecule, 
i.e. residues in contact, undergo motions with positive correlation, 
as one would expect for contact-driven motions.

One can identify three main blocks in the covariance plot~(fig. \ref{fig:covmat}):
the first one contains helices A, B, C, E, the loops between them and the EF loop;
the second one includes clearly the preF-loop, heme bound helix F, as well as the
first part of helix G, while the third block hosts the major part of helix
G and the C-terminal side of helix H.

Most residues in the first block, especially the ones belonging to
helical regions A, B and E (distal side), show a remarkable
anti-correlation with residues
localized at the beginning of the second block, belonging to the
proximal helix F and to helix G; in the third block the last turns of
helix H is bent at the C-terminal to allow closer
contacts with heme~\cite{pesce}.

\begin{figure}[htbp]
\begin{center}
  \includegraphics[width=0.5\textwidth]{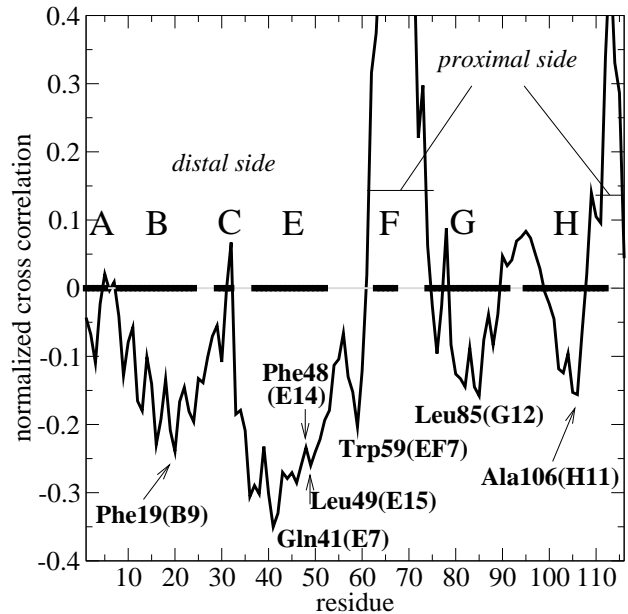}
\end{center}
\caption{
  \red{
  Normalized cross correlations between the proximal histidine
  F8 and the rest of the protein ($c_{68j}$, hence the peak raising to
  1.0 at $j = 68$): residues displaying significant anti-correlations
  with His68 are labeled in the plot.
  Like Phe19 (B9), they are strongly conserved throughout the trHb family~\cite{pesce},
  being relevant to prevent solvent access to the heme pocket (E14,
  EF7~\cite{pesce}), to stabilize heme bound ligand (E7~\cite{pesce})
  and to build the gate between the heme pocket and an apolar cavity running inside
  the protein matrix (G12, H11~\cite{milani1}).}
  }
\label{fig:hisf8}
\end{figure}

\red{
This division in domain of motions is similar to the one found
in~\cite{senoy1} for deoxymyoglobin, provided that one notes
the effect of the bending of C-terminal side in helix H, which
implies a correlated motion with the proximal side, as suggested by
fig.~\ref{fig:hisf8}, where normalized correlations between His68 (F8)
and the rest of the protein are shown.
Here the crucial role of small helix F in the dynamics
of the protein is underlined, since it contains the proximal
histidine, and the division of the protein in domains of motion
as described above is made more evident.
}

The covariance minima in the plot of figure~\ref{fig:hisf8} 
are particurlarly meaningful, being found between His F8 and other 
key residues of the protein.
Phe19 (B9), which has a bulky side chain, is responsible for the screening of
the distal cavity from the acqueous environment outside the molecule,
and is strongly conserved among trHbs;
in a position occupied by the distal histidine in vertebrate Hbs we
found Gln41 (E7), hydrogen bonded to Tyr20 (B10), which 
contribute to stabilize the heme-bound ligand~\cite{pesce}
and form a hydrogen bonding network in the heme
pocket, which is believed to be responsible for the different ligand
rebinding kinetics displayed by PtrHb and CtrHb in comparison 
with Mbs and Hbs~\cite{samuni}.

His68, taken here as a representative to deduce 
the motion of the whole proximal side from the covariance and
correlation plots (figg.~\ref{fig:covmat},~\ref{fig:hisf8}),
anti-correlates with hydrophobic Phe33 (CD1) as well, 
another strongly conserved residue among trHbs: 
together with the previous three residues they line precisely 
the distal cavity facing the heme group.
The anti-correlation of the distal and proximal sides is a clear
sign of the concerted motion which may allow the heme pocket to expand, 
thus making easier access to heme for ligands coming from
the apolar cavity that links the inner part of the protein to the solvent~\cite{pesce}, 
escaping the steric hindrance of the distal side residues.

\color{red}
Strong anti-correlation with the proximal histidine 
are displayed by Leu49 (E15), Leu85 (G12) and Ala/Val106 (H11) as well:
these residues lie at the bottom of the distal cavity, 
at the interface between the tunnel running inside the protein matrix
and the heme pocket~\cite{milani1, crespo}.

\begin{figure}[htbp]
\begin{center}
  \includegraphics[width=0.5\textwidth]{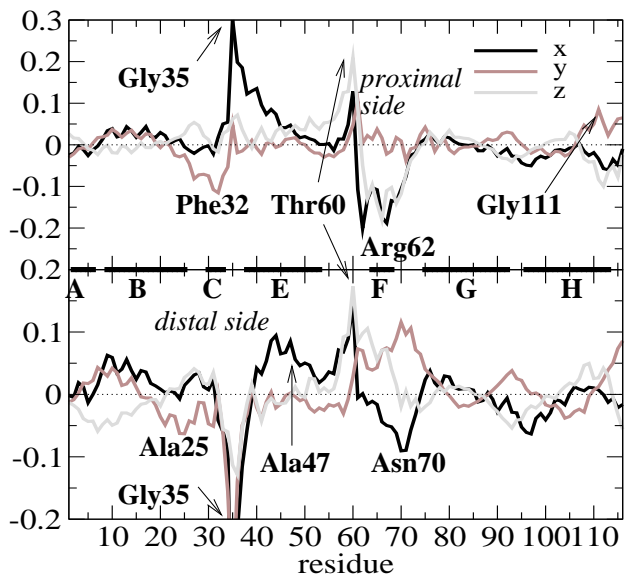}
\end{center}
\caption{
  \red{
  Components of normalized eigenvectors for the first two slowest
  modes of motion (1, top; 2, bottom; 
  ratio of corresponding eigenvalues: 1.16), 
  which bring a similar contribute to the dominant
  opening mechanism of the distal cavity, driven by the anticorrelated
  motions of the proximal (pre-F loop, helix F, loop FG and last
  part of helix H) and distal sides (helix C, CD loop and helix E
  especially). Residues with bulky side chains, strongly conserved in
  the family of trHbs and belonging to the hydrophic cluster
  preventing solvent access to the heme pocket~\cite{pesce}  
  are spatially located near the residues with biggest
  components, highlighted in the plot:
  Phe33 (CD1), Trp59 (EF7), Phe48 (E14). 
  The latter acts as gating residue in trHbN 
  from \emph{Mycobacterium Tuberculosis}~\cite{crespo}.
  }
}
\label{fig:12}
\end{figure}
The anticorrelated motion of the proximal and distal sides 
is made more visible by inspection of the components 
of the eigenvectors corresponding to the first two
slowest overdamped modes, plotted in figure~\ref{fig:12}.

Residues displaying the biggest deviations from their native positions
are highlighted: they belong to loops between helices lining the heme pocket
(CD and EF loops, pre-F region), and to helices enclosing the distal
and proximal sides (helix B and E, helix F and H).
These modes contribute substantially to the opening and closing of the distal
side, in agreement with previous studies on globins~\cite{senoy1, senoy2}. 

\begin{figure}[htbp]
\begin{center}
  \includegraphics[width=\textwidth]{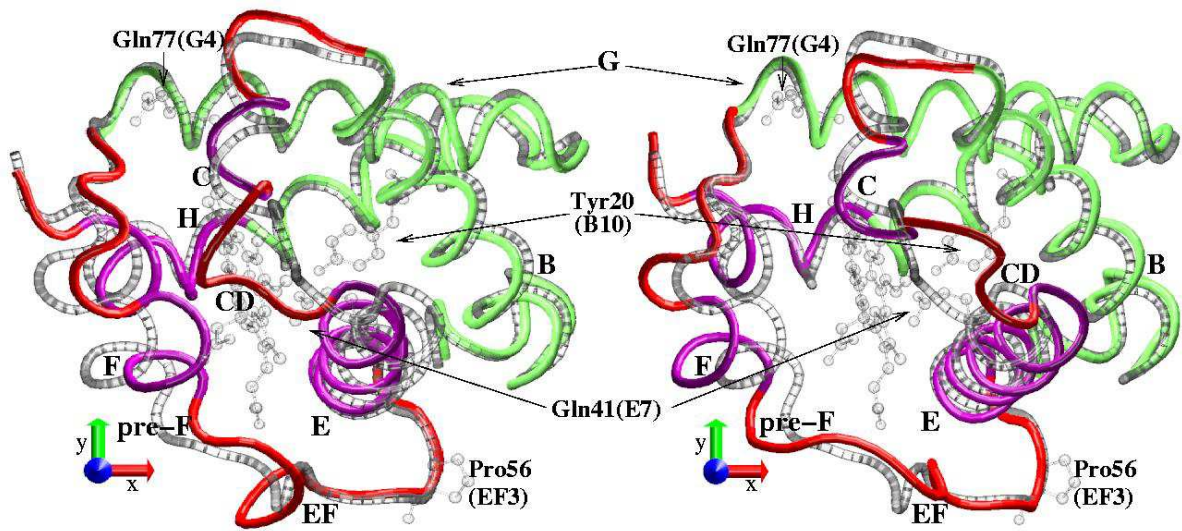}
\end{center}
\caption{
\red{
Open (left) and closed (right) 
conformations of the distal cavity, 
obtained by adding or subtracting 
the rescaled eigenvector of the first slowest mode
to the native positions of $\a$ carbons 
(scaling factor: 20).
Most mobile regions in the first mode are coloured in red (loops) and
purple (helices). 
Heme group and native structure are drawn in gray, as well as heme
bound ligand stabilizing residues Tyr B10 and Gln E7 and hinges of the distal side
opening mechanism - Pro EF3 and Gln G4
(figure drawn using VMD~\cite{vmd} and Raster3d~\cite{render}).
}
}
\label{fig:1ce}
\end{figure}
A detailed view of the conformations visited by the first mode is
shown in figure~\ref{fig:1ce}, where the open and closed structures of
the distal cavity are displayed, along with distal residues Tyr B10 and Gln E7.

From the covariance plot (fig.~\ref{fig:covmat})
and the component along the y axis of the second slowest
eigenvector of figure~\ref{fig:12} 
(although small, due to the normalization, which
enhances most mobile regions like loops)
one can notice the anticorrelation of the proximal histidine with the
residues identified to line the passage leading to the heme
pocket from the tunnel inside the protein 
(mainly Phe19, Leu85 and Ala106, already evidenced in
fig.~\ref{fig:hisf8})~\cite{milani1, bolognesi, milani2, crespo}.
The anti-correlation between the two groups of residues
hints at a possible mechanism for the passage of ligands to the heme
pocket, through the enlargement of the gate: 
the presence of the apolar cavity has been proposed 
to contribute effectively to the fast rebinding of ligands on heme, 
together with the hydrogen bonding network 
in the distal side, as already pointed out~\cite{milani1, samuni,
  milani2, bolognesi, milani3}.

\begin{figure}[htbp]
\begin{center}
  \includegraphics[width=0.50\textwidth]{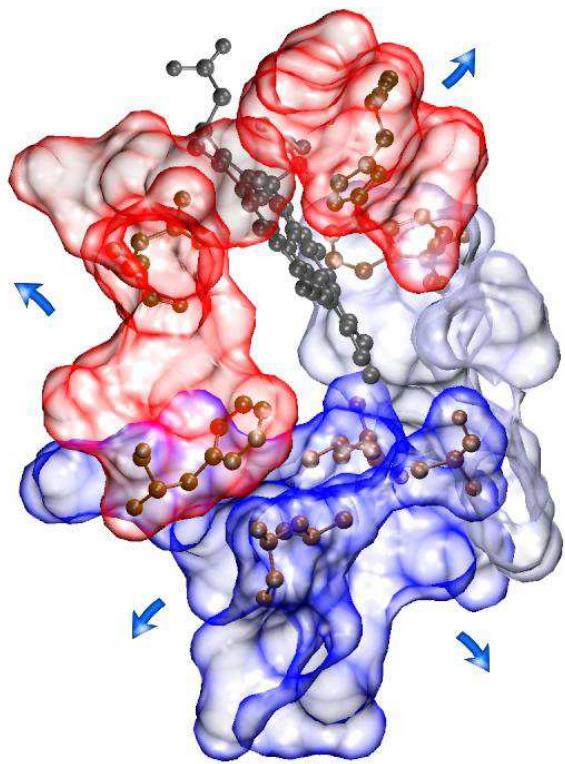}
\end{center}
\caption{
  Schematic representation of near equilibrium motions of groups of
  residues delimiting the heme pocket, inferred from covariance
  analysis: solvent accessible surfaces of residues delimiting the 
  apolar cavity (res. 6, 12, 16, 17, 49, 53, 85, 89, lower left), 
  and the distal cavity (proximal side: res. 64, 68, 71, upper right; 
  distal side: res. 19, 20, 32, 33, 41, upper left) 
  are shown with a 1.4 \AA \: radius probe.
  Ball-and-stick representation is used for His68 and the residues
  labeled in figure~\ref{fig:hisf8}.
  The cluster in the lower right (res. 48, 51, 52, 59, 105, 109) defines a
  narrower cavity~\cite{milani2}. 
  Figure drawn with VMD~\cite{vmd}, rendered with Raster3d~\cite{render}.
}
\label{fig:cavity}
\end{figure}

The combined motion of the main blocks is compatible with a pumping 
mechanism: according to the results obtained in this study, the native
state conformation of the two truncated hemoglobins is such that small
displacements of the atoms, due to stochastic
interactions with the solvent, produce an anti-correlated motion 
of the proximal and distal sides, which line the heme pocket,
bringing atoms back to equilibrum positions.
These movements may facilitate the diffusion of
small ligands such as $O_2$ and $CO$ to heme through the protein tunnel,
exploiting its volume variations~\cite{crespo}.

\subsection{Elasticity and time scale}

An estimate of the elastic constant $\g$ of the model can be computed
fitting the experimental temperature factors of the X-ray structures
with the theoretical ones, obtained from the mean square displacements
of $C^\a$s, according to equation (\ref{eqn:B-factors}).  Following
the method used in~\cite{bahar1} to fit the data (i.e. by matching the
areas of the surface enclosed by the two data sets) and averaging the
values found for the two proteins yields $\g = 0.20\:Nm^{-1}$, with
a tolerance of $0.05\:Nm^{-1}$ between averaged values (the
introduction of heme in the network of interactions leads to a decreas
of the value of the elastic constant, since it enhances the local connectivity of
the buried residues in the heme pocket).  The order of magnitude
obtained for $\g$ agrees with estimated values for the elastic
constant of single parameter models~\cite{tirion, atilgan, mich,
neri}.

The importance of friction due to the solvent in determining the rates of 
functional motions of proteins, as it slows down the relaxation times of 
large-scale displacements predicted by normal mode analysis,
has recently been underlined~\cite{ma}:
in the framework of the Langevin dynamics introduced with
equation~(\ref{eqn:langevin}), we estimate order of magnitude for the reference
decay time $\tau$ of the first two modes of motion previously
described, through an effective value for the friction
coeffient $f$, chosen to be the same for all residues for simplicity. 
A lower limit for $f$ is the value computed from an all-atom
simulation in~\cite{karplus}, whereas here whole residues are considered
(although the effective radii associated with such an estimate are bigger 
than the Van der Waals radii of the atoms in the simulation, 
hinting at a collective character of the simulated 
displacements~\cite{karplus}, the motions predicted 
by the slowest modes involve many more residues in distant parts 
of the protein and a larger value for the friction may be expected).
As an upper limit, the friction relative to the whole proteins 
(both PtrHb and CtrHb roughly fit a cubic box of side $3.5\,nm$)
moving in water at physiological conditions is calculated 
from Stoke's law (see~\cite{howard}, chapter 3).
We obtained $f\,\sim\,4\,\div\,70\,pN\,m^{-1}\,s$
(similar ranges for the values of friction coefficients have been 
extracted from molecular dynamics simulations~\cite{hinsen}).

The corresponding reference relaxation time $\tau$ in 
the Langevin dynamics of equations~\ref{eqn:langevin} 
and~\ref{eqn:time-correlation}
lies within the range $0.02\,\div\,0.35\,ns$, while 
the relaxation time associated with eigenmode $i$ will be
$\tau_i = \frac{\tau}{\l_i}$ (where $\l_i$ is the eigenvalue relative
to that eigenmode): the two slowest eigenmodes display 
relaxation times for the related motions approximately 
within the range $0.2\,\div\,3.5\,ns$.

This range of time scale is compatible with CO rebinding kinetics
of Mbs and Hbs, while PtrHb and CtrHb behave quite
differently~\cite{samuni}: the explanation proposed for the different behaviour 
relies on the hydrogen-bonding network formed in the distal
cavity of these trHbs, which is absent in invertebrate globins and is
beyond the possibility of the simple model used here, 
which underlines instead the common characteristics of globins and
trHbs.


\color{black}
\section{Conclusions}

It has been shown how a simple coarse-grained approach can bring insights
into the functional motions of two small proteins of the truncated hemoglobins family,
PtrHb and CtrHb, near equilibrium vibrational properties of the structures
modeled as a gaussian network of interacting $\a$ carbons and $\b$ centroids.  

The key point in the analysis performed here is the information
extracted from the covariance matrix in its reduced form
and from the two slowest modes of fluctuation: 
negative correlations between residues set far apart 
in the tridimensional structure are particularly useful, being non
trivial and hinting at the collective character of the motions.

This information has been used in the present work 
to confirm within such a simplified approach 
the mechanism which is believed to facilitate small ligands
diffusion to the heme pocket and the iron atom.
The cavity delimited by several key hydrophobic residues, 
providing a path from the surface of the protein to the heme pocket
~\cite{pesce, milani1, milani3, bolognesi},, 
is able to enlarge its volume allowing the passage 
of small molecules to the distal side~\cite{samuni, crespo}, 
as it is inferred from the anti-correlations between 
the displacements of the opposite sides of the heme pocket.

Excitations, due to interactions
between the molecule and the solvent, produce deviations from
equilibrium followed by a decay towards the native state.
The collective behaviour of the return back to equilibrium, 
produced by a superposition of overdamped motions, 
allow the volume of the inner cavities to vary accordingly.

Through a fit of the mean square displacements of $\a$ carbons from their minimum energy
configuration with the experimental temperature factors for the two
structures under study, a rough estimate of the order of magnitude of
time scale for functionally relevant motions has
been given, in reasonable agreement with known properties of globular proteins. 

This suggests the validity of the simple gaussian approach as a means to get a
fast picture of the near-native functional motions of globular
proteins, yet in agreement with the results obtained using more
accurate and computationally demanding tools.

The description given by the simple model used here does not provide
atomic details, keeping the analysis at a coarse-grained level.
Still the use of the effective $\b$ centroid for each residue, 
along with the $C^\a$, helps in characterizing with more
adherence to reality the displacements of residues side-chains, thus
getting a closer agreement with more detailed approaches.

\section{Acknowledgments}
I wish to thank Amos Maritan and Gianluca Lattanzi for  
the invaluable suggestions and the critical reading of the manuscript.
The collaboration with Cristian Micheletti in the initial stage of
this work is gratefully acknowledged.

\bibliography{trHb}
\bibstyle{apsrev}

\end{document}